\documentclass[12pt]{article}
\usepackage{epsfig}
\usepackage{psfrag}
\usepackage{latexsym}
\usepackage{indentfirst}
\usepackage{fancyhdr}
\usepackage{amssymb}
\usepackage{amsmath}
\usepackage{amsfonts}
\usepackage{cite}
\usepackage{bbold}
\usepackage{color}
\usepackage[footnotesize]{caption2}
\usepackage{graphicx}
\usepackage[center,footnotesize,hang]{subfigure}
\usepackage{url}
\usepackage{bm}
\usepackage{slashed}

\makeatletter

\newcommand{\Rmnum}[1]{\expandafter\@slowromancap\romannumeral #1@}
\makeatother

\newcommand{\pa}{\partial}

\def\be{\begin{equation}}
\def\ee{\end{equation}}
\def\bc{\begin{center}}
\def\ec{\end{center}}
\def\bea{\begin{eqnarray}}
\def\eea{\end{eqnarray}}
\def\la{\label}

\def\nn{\nonumber}

\catcode`@=11
\def\marginnote#1{}
\newcount\hour
\newcount\minute
\newtoks\amorpm
\hour=\time\divide\hour by60
\minute=\time{\multiply\hour by60 \global\advance\minute by-\hour}
\edef\standardtime{{\ifnum\hour<12 \global\amorpm={am}%
        \else\global\amorpm={pm}\advance\hour by-12 \fi
        \ifnum\hour=0 \hour=12 \fi
        \number\hour:\ifnum\minute<10 0\fi\number\minute\the\amorpm}}
\edef\militarytime{\number\hour:\ifnum\minute<10 0\fi\number\minute}
\def\draftlabel#1{{\@bsphack\if@filesw {\let\thepage\relax
   \xdef\@gtempa{\write\@auxout{\string
      \newlabel{#1}{{\@currentlabel}{\thepage}}}}}\@gtempa
   \if@nobreak \ifvmode\nobreak\fi\fi\fi\@esphack}
        \gdef\@eqnlabel{#1}}
\def\@eqnlabel{}
\def\@vacuum{}
\def\draftmarginnote#1{\marginpar{\raggedright\scriptsize\tt#1}}
\def\draft{\oddsidemargin 0.0truein
        \def\@oddfoot{\sl preliminary draft \hfil
        \rm\thepage\hfil\sl\today\quad\militarytime}
        \let\@evenfoot\@oddfoot \overfullrule 3pt
        \let\label=\draftlabel
        \let\marginnote=\draftmarginnote
   \def\@eqnnum{(\theequation)\rlap{\kern\marginparsep\tt\@eqnlabel}%
\global\let\@eqnlabel\@vacuum}  }
\catcode`@=12
\begin{document}
\title{\bf {Killing Vectors in Spacetime of the De Sitter Invariant Special Relativity}}

\author{Mu-Lin Yan\footnote{E-mail address: mlyan@ustc.edu.cn}
\\Interdisciplinary Center for Theoretical Study \\  Department of Modern Physics\\
University of Science and Technology of China\\ Hefei, Anhui 230026, China
}
\maketitle

\abstract{In this paper, we use the Killing vector method to formulate the de Sitter/Anti-de Sitter invariant special relativity (dS/AdS-SR). Through solving the Einstein equation with $\Lambda\neq 0$, the basic inertial metric for dS/AdS-SR is determined to be the Beltrami metric $B_{\mu\nu}(x)$. The corresponding Killing equations are system of ten simultaneous partial differential equations of first order. Their most general solutions were obtained, and all the ten independent Killing vectors were found out. These results confirm that the Beltrami metric has maximal spacetime symmetry. The ten Killing-Noether charges are obtained. They are energy, momenta, Lorentz boost and angular momentum in SR-theory with $\Lambda\neq 0$. Consequently, dS/AdS-SR is consistently established for the vacuum with $\Lambda\neq 0$ via Killing vector method rather than the unpopular classical domain theory.
}

\section{Introduction}\label{sec:intr}
Common Special Relativity (SR) is invariant under Poincar\'e transformations and its basic space-time metric is Minkowski metric $g_{\mu\nu}=\eta_{\mu\nu}\equiv {\rm{diag}}\{+,-,-,-\}$, which satisfies the vacuum (or empty spacetime) Einstein equation without universal Einstein Cosmologic Constant (ECC) $\Lambda$. It is easy to see when $\Lambda \neq 0$, the Minkowski metric  will no longer be a solution of the vacuum Einstein equation because a new term $\Lambda g_{\mu\nu}$ will emerge in the equation. In this case the common SR should naturally become the de Sitter ( or Anti de Sitter) invariant Special Relativity (dS/AdS-SR)~\cite{Dirac35,Lu74}(see also~\cite{yan2} and references within). In other words, the de Sitter/Anti de Sitter invariant Special Relativity is the Special  Relativity in the vacuum spacetime with the non-zero universal Einstein Cosmology Constant $\Lambda$. It is essential that the vacuum of dS/AdS-SR is different from one of common SR.

About the end of last century, the accelerating expansions of the Universe were discovered~\cite{Riess,Perl}. The accelerating expansions of the Universe indicate that there is an  effective positive cosmologic constant $\Lambda_{\rm eff}\equiv \Lambda+8\pi G\rho_{\rm dark\;energy}\neq 0$ in the real world spacetime, where $\Lambda$ is ECC that is a universal constant, $G$ is Newton gravitational constant and $\rho_{\rm dark\;energy}$ is density of dark energies (e.g., see~\cite{review,review1,yan3}). We address that the fact of $\Lambda_{eff}\neq 0$ in general means that ECC $\Lambda\neq 0$ and $\rho_{\rm dark\; energy}\neq 0$. It is {\it ad hoc} to assume $\Lambda=0$ artificially in the studies of the cosmology of the time. Thus, after this discover, the theory of dS/AdS-SR attracts great interests~\cite{yan2,yan3,Guo2,Guo2-1,Guo2-2,Guo1,Guo1-1,Guo1-2,Lu05,yan1,cxy,fy,zhao,ru,yan4,yan5}.

The SR basic spacetime metric $g_{\mu\nu}$ plays a pivotal role in SR-theory, which can be determined by following requirements:
\begin{enumerate}
\item{ The $g_{\mu\nu}$ satisfies the vacuum Einstein equation.}
\item{ In the $g_{\mu\nu}$ spacetime, the motions of free particles are inertial.}
\item{$g_{\mu\nu}$ has maximal spacetime symmetry.}
\end{enumerate}
In addition, it should be also kept in mind that the non-relativistic limit of any relativistic mechanics has to be the common Newtonian mechanics~\cite{Landau}. When ECC $\Lambda=0$, the solution satisfying the above three requirements is $g_{\mu\nu}=\eta_{\mu\nu}$. When $\Lambda\neq 0$, the situation becomes non-trivial and  challenging. For this case, based on analysis of symmetrical space by using classical domain theory method~\cite{hua2}, a remarkable metric $g_{\mu\nu}(\lambda,x)$ was suggested in Ref.~\cite{Lu74} (where $\lambda$ is a constant), which satisfies the $2^{\rm nd}$
 requirement (i.e., inertial motion law for free particles holed in the spacetime with this $g_{\mu\nu}(\lambda,x)$). In this present paper, we take this classical domain spacetime metric $g_{\mu\nu}(\lambda,x)$ to be an ansatz for solving the vacuum Einstein equation with $\Lambda\neq 0$. We will find that $g_{\mu\nu}(\lambda, x)$ with $\lambda=\Lambda/3$ (see Eq.~(\ref{34}) below) is the solution, i.e., $g_{\mu\nu}(\lambda,x)|_{\lambda=\Lambda/3}\equiv B_{\mu\nu}(x)$ is the solution of the vacuum Einstein equation with $\Lambda\neq 0$, and it will be called Beltrami metric. Hence, $B_{\mu\nu}(x)$  satisfies both the $1^{\rm st}$ requirement and $2^{\rm nd}$ requirement. Next we should examine whether $B_{\mu\nu}(x)$ satisfies the $3^{\rm rd}$ requirement. This is the main aim of the present paper. We will present explicit calculations to solve the Killing vector equation of $B_{\mu\nu}(x)$, which is a system of ten simultaneous partial differential equations of first order (e.g., see the 13th chapter in~\cite{Weinberg72}), and all the corresponding  Killing-Neother charges are found out.

The rest of the paper is organized as follows.
In section II, we solve the vacuum Einstein equation with $\Lambda\neq 0$. In this way, we find that the basic metric of dS/AdS-SR is the Beltrami metric. In section III, we solve the Killing equation of Beltrami metric. It is a system of ten simultaneous partial differential equations of first order. All Killing vectors in the Beltrami spacetime are found out. Section IV devotes to calculating Killing-Noether charges, and confirms the metric $B_{\mu\nu}(x)$ has maximal spacetime symmetry. Finally, we briefly summarize and discuss our results in this paper.

\section{Basic metric of dS/AdS-invariant special relativity}

The Einstein equation with cosmologic constant is given by,
\bea\la{be}
\mathcal{R}_{\mu\nu}-{1\over 2} g_{\mu\nu} \mathcal{R}+ \Lambda g_{\mu\nu}=0,
\eea
where $\mathcal{R}_{\mu\nu}$ and $\mathcal{R}$ are the Ricci tensor and curvature scalar of 4-dimensional Riemann geometry respectively. In~\cite{Lu74}, the following metric ansatz was suggested,
\bea \la{b74}
g_{\mu\nu}(\lambda,x)= {\eta_{\mu\nu}\over \sigma(\lambda,x)}+{\lambda \eta_{\mu\alpha}\eta_{\nu\beta}x^\alpha x^\beta\over  \sigma(\lambda,x)^2},~~~{\rm with}~~~\sigma(\lambda,x)=1-{\lambda}{\eta_{\mu\nu}x^\mu x^\nu}\,,
\eea
where $\lambda$ is a constant. Straightforward geometry calculations give us
\bea\la{33-5}
&& g^{\mu\nu}(\lambda,x)=\sigma(\lambda,x)(\eta^{\mu\nu}-\lambda x^\mu x^\nu),\\
\la{33-6} &&\Gamma^\rho_{\mu\nu}(\lambda,x)={\lambda\over \sigma(\lambda,x)}(\delta^\rho_\mu \eta_{\nu\lambda}x^\lambda
+\delta^\rho_\nu \eta_{\mu\lambda}x^\lambda), \\
\la{33-7} &&\mathcal{R}^\rho_{\;\;\lambda\mu\nu}(\lambda,x)=\lambda \left[g_{\lambda\mu}(\lambda,x)\delta^\rho_\nu-g_{\lambda\nu}(\lambda,x) \delta^\rho_\mu\right],\\
\la{33-8}&& \mathcal{R}_{\mu\nu}(\lambda,x)={3\lambda }g_{\mu\nu}(\lambda,x),\\
\la{33-9}&& \mathcal{R}(\lambda)={12\lambda}={\rm constant.}
\eea
Substituting Eqs.~(\ref{33-8}, \ref{33-9}) into Eq.~\eqref{be}, we obtain
\bea\la{34}
\lambda={\Lambda\over 3}.
\eea
Consequently, the solution of the vacuum Einstein equation is
\bea\la{sol}
g_{\mu\nu}(x)\equiv B_{\mu\nu}(x)={\eta_{\mu\nu}\over \sigma(x)}+{\Lambda \eta_{\mu\alpha}\eta_{\nu\beta}x^\alpha x^\beta\over 3 \sigma(x)^2},~~{\rm with}~~\sigma(x)\equiv\sigma(\lambda,x)|_{\lambda=\Lambda/3}=1-{\Lambda\over 3}{\eta_{\mu\nu}x^\mu x^\nu}.
\eea
We call $B_{\mu\nu}(x)$ the Beltrami metric, and hence the $1^{\rm st}$ requirement listed in the last section is satisfied. The metric $g_{\mu\nu}(x)$ which is the solution to Eq.~\eqref{be} have dS/AdS-spacetime symmetry~\cite{Dirac35}.

The inertial motion law for free particle in the Beltrami spacetime ${\bf \cal{B}}$ described by $B_{\mu\nu}(x)$ has been discussed in~\cite{Lu74,yan2,yan1}. In order to clarify the notations which will be used below, we shall recapitulate the key points here. The inertial motion law in ${\bf \cal{B}}$ requires that the free particles in ${\bf \cal{B}}$ move uniformly along the straight line (or geodesic). Namely, by means of the principle of least action (which is the equivalent of the equation of motion along geodesic line in ${\bf \cal{B}}$ ):
\bea\la{b-1}
\delta S\equiv \delta\left[-mc\int ds\right]=-mc\;\delta\int\sqrt{B_{\mu\nu}(x)dx^\mu dx^\nu}=0\,,
\eea
we can get the solution as follows,
\bea\la{b-2}
\ddot{\bm{x}}=0,\;\;{\rm or}~\;\bm{v}=\dot{\bm{x}}={\rm constant},
\eea
where $S=-mc\int ds$ is Landau-Lifshitz action for free particle~\cite{Landau}£¬and $\dot{\bm{x}}$ and
$\ddot{\bm{x}}$ are the velocity and the acceleration respectively. The solution of Eq.~\eqref{b-2} have been obtained by solving Eq.~\eqref{b-1} in Refs.~\cite{yan2,yan1}, and hence the $2^{nd}$ requirement listed in the last section is satisfied in the Beltrami spacetime ${\bf \cal{B}}$. From the Landau-Lifshitz action $S=\int L dt$ in ${\bf \cal{B}}$-spacetime, we have
\bea\la{b-3}
L=-mc{ds\over dt}=-mc{\sqrt{B_{\mu\nu}(x)dx^\mu dx^\nu}\over dt}=-mc{\sqrt{B_{\mu\nu}(x)\dot{x}^\mu \dot{x}^\nu}}.
\eea
Substituting (\ref{sol}) into (\ref{b-3}) gives
\bea\la{b-4}
L=-mc^2  \sqrt{9 (c^2- \dot{\bm{x}}^2)+3\Lambda[-\bm{x}^2\dot{\bm{x}}^2+(\bm{x}\cdot\dot{\bm{x}})^2
+c^2(\bm{x}-\dot{\bm{x}}t)^2 ] \over c^2 [3+\Lambda(\bm{x}^2-c^2t^2)]^2}.
\eea
It is easy to see when $\Lambda\rightarrow 0$£¬we have
\bea\la{b2-5}
L\longrightarrow L_{Eins}=-mc^2\sqrt{1-{\bm{\dot{x}}\over c^2}^2},
\eea
where $L_{Eins}$ is well known Lagrangian of common SR which is Pioncar\'e invariant~\cite{Landau}.
By using the Euler-Lagrangian equation
\bea\la{3-27}
{d \over dt} {\pa L \over \pa \dot{\bm{x}}}-{\pa L \over \pa \bm{x}}=0,
\eea
and noting $L=L(t, \bm{x}, \dot{\bm{x}})$, $\pa /\pa \bm{x}\equiv \nabla=(\pa /\pa x^1)\bm{i}+(\pa /\pa x^2)\bm{j}+(\pa /\pa x^3)\bm{k}$, we can also obtain Eq.~\eqref{b-2}. The calculations are straightforward and non-trivial~\cite{yan2,yan1}.

In following section we will focus on the $3^{\rm rd}$ requirement for basic metric of SR. We shall solve the Kiling vector equation to examine whether $B_{\mu\nu}(x)$ has maximal symmetry or not.

\section{Killing vectors in Beltrami spacetime }

In order to understanding the geometry of ${\bf \cal{B}}$, and further to reveal the conservation laws in the mechanics of dS/AdS-SR, we derive the Killing vectors in this section. The metric in the Betrami spacetime ${\bf \cal{B}}$ is $g_{\mu\nu}(x)=B_{\mu\nu}(x)$. Considering a infinitely small coordinate transformation:
\bea\la{b3-1}
x^\mu\longrightarrow x'^{\mu}=x^\mu+\epsilon \xi^\mu(x),~~\text{with}~~|\epsilon|<<1\,,
\eea
where $\xi^\mu(x)$ is generators of the transformation, the condition that $g_{\mu\nu}(x)$ is invariant under this transformation is given by
\bea\la{b3-2}
\mathcal{L}_\xi g_{\mu\nu}(x)=0,
\eea
where $\mathcal{L}_\xi g_{\mu\nu}(x)$ is the Lee derivative of $g_{\mu\nu}(x)$, and then $\xi^\mu(x)$ is the Killing vector. Hence $\xi^\mu (x)$ is determined by following Killing vector equation (see, e.g., the 13th chapter of~\cite{Weinberg72}):
\bea\la{b3-3}
\xi_{\mu;\nu}+\xi_{\nu;\mu}=0\,,
\eea
where the covariant derivative $\xi_{\mu;\nu}=\xi_{\mu,\nu}-\Gamma^\lambda_{\mu\nu}\xi_\lambda$.
For all possible values of $\mu$ and $\nu$, the Killing equation of Eq.~\eqref{b3-3} reads as
\begin{eqnarray}
&&\la{bk0_add} {\pa\xi_0\over \pa x^0}={2\Lambda x^0\over 3\sigma(x)}\xi_0,\\
\la{bk1} && {\pa\xi_1\over \pa x^1}={-2\Lambda x^1\over 3\sigma(x)}\xi_1,\\
\la{bk2} && {\pa\xi_2\over \pa x^2}={-2\Lambda x^2\over 3\sigma(x)}\xi_2,\\
\la{bk3} && {\pa\xi_3\over \pa x^3}={-2\Lambda x^3\over 3\sigma(x)}\xi_3,\\
\la{bk4}
&& {\pa\xi_0\over \pa x^1}+{\pa\xi_1\over \pa x^0}={2\Lambda\over3\sigma(x)}(-\xi_0x^1+\xi_1x^0)\,,\\
\la{bk5}&& {\pa\xi_0\over \pa x^2}+{\pa\xi_2\over \pa x^0}={2\Lambda\over3\sigma(x)}(-\xi_0x^2+\xi_2x^0)\,,\\
\la{bk6}&& {\pa\xi_0\over \pa x^3}+{\pa\xi_3\over \pa x^0}={2\Lambda\over3\sigma(x)}(-\xi_0x^3+\xi_3x^0)\,,\\
\la{bk7}&&{\pa\xi_1\over \pa x^2}+{\pa\xi_2\over \pa x^1}={-2\Lambda\over3\sigma(x)}(\xi_2x^1+\xi_1x^2)\,,\\
\la{bk8}&&{\pa\xi_1\over \pa x^3}+{\pa\xi_3\over \pa x^1}={-2\Lambda\over3\sigma(x)}(\xi_3x^1+\xi_1x^3)\,,\\
\la{bk9}&&{\pa\xi_2\over \pa x^3}+{\pa\xi_3\over \pa x^2}={-2\Lambda\over3\sigma(x)}(\xi_2x^3+\xi_3x^2)\,,
\end{eqnarray}
where $\sigma(x)=1-{\Lambda\over 3}{\eta_{\mu\nu}x^\mu x^\nu}$. Our purpose
is to solve the above ten simultaneous partial differential equations of first order. For convenience and notation compactness, we introduce the notations $f(x^1,x^2,x^3)\equiv f(\slashed{0})$, $f(x^0,x^2,x^3)\equiv f(\slashed{1})$, $f(x^0,x^1,x^3)\equiv f(\slashed{2})$ and $f(x^0,x^1,x^2)\equiv f(\slashed{3})$. That is to say,  $f(\slashed{\mu})$ is a multivariable function of $x$, but is independent of the $\mu$-th component $x^{\mu}$. From Eq.~\eqref{bk0_add}, we can obtain
\bea\nn
{d\xi_0\over \xi_0}={2\Lambda x^0dx^0\over 3(1-{\Lambda\over 3}\eta_{\mu\nu}x^\mu x^\nu)}
={\Lambda d[(x^0)^2]\over 3-\Lambda((x^0)^2-{\bm{x}}^2)}\,.
\eea
Performing integrals on both sides of the above equation, we have
\bea
\label{eq:xi0}
\ln \xi_0=\ln\left[{c(\slashed{0})\over \sigma(x)}\right]\Rightarrow \xi_0={c(\slashed{0})\over \sigma(x)}\,.
\eea
In a similar way, from the equations (\ref{bk1}), (\ref{bk2}), (\ref{bk3}), the following relations can be obtained,
\bea
\label{eq:xi123}
\xi_1={c(\slashed{1})\over \sigma(x)},\;\;\xi_2={c(\slashed{2})\over \sigma(x)}, \;\;\xi_3={c(\slashed{3})\over \sigma(x)},
\eea
Substituting Eqs.~(\ref{eq:xi0}, \ref{eq:xi123}) into (\ref{bk4})--(\ref{bk9}), we have
\bea\la{bc1}
&&{\pa c(\slashed{1})\over \pa x^2}+{\pa c(\slashed{2})\over \pa x^1}=0\,,\\
\la{bc2}
&&{\pa c(\slashed{1})\over \pa x^3}+{\pa c(\slashed{3})\over \pa x^1}=0\,,\\
\la{bc3}
&&{\pa c(\slashed{1})\over \pa x^0}+{\pa c(\slashed{0})\over \pa x^1}=0\,,\\
\la{bc4}
&&{\pa c(\slashed{2})\over \pa x^3}+{\pa c(\slashed{3})\over \pa x^2}=0\,,\\
\la{bc5}
&&{\pa c(\slashed{2})\over \pa x^0}+{\pa c(\slashed{0})\over \pa x^2}=0\,,\\
\la{bc6}
&&{\pa c(\slashed{3})\over \pa x^0}+{\pa c(\slashed{0})\over \pa x^3}=0\,.
\eea
As a consequence, we see that $\frac{\partial c(\slashed{i})}{\partial x^{j}}$ is independent of both $x^{i}$ and $x^{j}$, and
\begin{equation}
\frac{\partial^{3}c(\slashed{0})}{\partial x^{1}\partial x^{2}\partial x^{3}},\quad \frac{\partial^{3}c(\slashed{1})}{\partial x^{0}\partial x^{2}\partial x^{3}},\quad \frac{\partial^{3}c(\slashed{2})}{\partial x^{0}\partial x^{1}\partial x^{3}},\quad \frac{\partial^{3}c(\slashed{3})}{\partial x^{0}\partial x^{1}\partial x^{2}}
\end{equation}
are constants. Hence the most general form of the function $c(\slashed{\mu})$ is as follows
\begin{equation}
\label{eq:expansion_solution}
\begin{aligned}
c(\slashed{0})&=a_0+b_{01}x^1+b_{02}x^2+b_{03}x^3+d_{03}x^{1}x^{2}+d_{02}x^{1}x^{3}+d_{01}x^{2}x^{3}+f_0x^{1}x^{2}x^{3}\,,\\
c(\slashed{1})&=a_1+b_{10}x^0+b_{12}x^2+b_{13}x^3+d_{13}x^{0}x^{2}+d_{12}x^{0}x^{3}+d_{10}x^{2}x^{3}+f_1x^{0}x^{2}x^{3}\,,\\
c(\slashed{2})&=a_2+b_{20}x^0+b_{21}x^1+b_{23}x^3+d_{23}x^{0}x^{1}+d_{21}x^{0}x^{3}+d_{20}x^{1}x^{3}+f_2x^{0}x^{1}x^{3}\,,\\
c(\slashed{3})&=a_3+b_{30}x^0+b_{31}x^1+b_{32}x^2+d_{32}x^{0}x^{1}+d_{31}x^{0}x^{2}+d_{30}x^{1}x^{2}+f_3x^{0}x^{1}x^{2}\,,
\end{aligned}
\end{equation}
where $a_i$, $b_{ij}$, $d_{ij}$ and $f_i$ with $i, j=0, 1, 2, 3$ are real. Inserting Eq.~\eqref{eq:expansion_solution} into Eqs.~(\ref{bc1}, \ref{bc2}, \ref{bc3}, \ref{bc4}, \ref{bc5}, \ref{bc6}), we obtain the following constraints
\begin{equation}
\label{eq:parameters_cons}
\begin{aligned}
&d_{01}=d_{02}=d_{03}=d_{10}=d_{12}=d_{13}=d_{20}=d_{21}=d_{23}=d_{30}=d_{31}=d_{32}=0\,,\\
&f_0=f_1=f_2=f_3=0,\quad b_{01}=-b_{10},\quad b_{02}=-b_{20},\quad b_{03}=-b_{30} \,,\\
&b_{12}=-b_{21},\quad b_{13}=-b_{31},\quad b_{23}=-b_{32}\,.
\end{aligned}
\end{equation}
Therefore the Killing vector of the Betrami metric is
\bea\la{bkkl}
\xi_\mu(x)={3c(\slashed{\mu})\over 3-\Lambda\eta_{\mu\nu}x^\mu x^\nu},
\eea
with
\bea\la{bcc}
\left(\begin{array}{c}
c(\slashed{0})\\
c(\slashed{1})\\
c(\slashed{2})\\
c(\slashed{3})
\end{array}\right)=\left(\begin{array}{lccr}
                    0& -b_{10} & -b_{20}& -b_{30}\\
                    b_{10}& 0& b_{12}& b_{13}\\
                    b_{20}& -b_{12}&0& b_{23}\\
                    b_{30}& -b_{13}& -b_{23}&0
                    \end{array}\right)
\left(\begin{array}{c}
x^0\\
x^1\\
x^2\\
x^3
\end{array}\right)+\left(\begin{array}{c}
a_0\\
a_1\\
a_2\\
a_3
\end{array}\right),
\eea
where $b_{\mu\nu}$ and $a_{\mu}$ are ten independent constants, and hence Eq.~\eqref{bkkl} indicates that there are ten independent Killing vectors in the Beltrami spacetime ${\bf \cal{B}}$ described by $g_{\mu\nu}(x)=B_{\mu\nu}(x)$. Noting the dimension of ${\bf \cal{B}}$ is $N=4$, and $N(N+1)/2=10$. Consequently the Beltrami metric $B_{\mu\nu}(x)$ has maximum spacetime symmetry (see, e.g., the 13th chapter of ~\cite{Weinberg72}), and then we have proved that the $3^{rd}$ requirement listed in the introduction section is satisfied for $B_{\mu\nu}(x)$. From (\ref{sol}), we can read off the contravariant metric tensor in ${\bf\cal{B}}$ as
\bea\la{sol-1}
B^{\mu\nu}(x)=\sigma(x)\left(\eta^{\mu\nu}-{\Lambda\over3}x^\mu x^\nu\right).
\eea
Consequently the contravariant Killing vector in ${\bf\mathcal{B}}$ is
\bea\la{bshh3}
\xi^\mu(x)=B^{\mu\nu}(x)\xi_\nu(x)=\eta^{\mu\nu}c(\slashed{\nu})-{\Lambda\over3} x^\mu x^\nu c(\slashed{\nu})\,,
\eea
where $\eta^{\mu\nu}c(\slashed{\nu})$ refers to $\sum_{\nu=0}^{3}\left(\eta^{\mu\nu}c(\slashed{\nu})\right)$.
Substituting (\ref{bshh3}) into (\ref{b3-1}) gives
\bea\la{b33-1}
x^\mu\longrightarrow x'^{\mu}=x^\mu+\epsilon \left(\eta^{\mu\nu} c(\nu\hskip-0.08in /)-{\Lambda\over 3}x^\mu x^\nu c(\slashed{\nu}) \right),~~ \rm{where}~~|\epsilon|<<1\,,
\eea
which is the infinitely small coordinate transformation preserved by the Beltrami metric. Hence we can conclude that the Betrami metric (\ref{sol}) fully satisfies the three requirements for the basic spacetime metric of SR claimed in the introduction section.

\section{\label{sec:sec:killing_noether_charges}Noether theorem and Killing-Noether charges}

For clarifying the notations we briefly review the well known Noether theorem (see, e.g.,~\cite{Noether,Arnold}) at first, and then we present detailed calculations for Killing-Noether charges in the following.

\noindent {\bf (A) Noether theorem}

Considering a mechanics system, its dynamical behaviors are described by the Lagrangian  $L(t, \bm{q}, \dot{\bm{q}})$ and the Euler-Lagrange equation arising from the variation $\delta \int  L(t, \bm{q}, \dot{\bm{q}})dt=0$. If the action $S\equiv \int L(t, \bm{q}, \dot{\bm{q}}) dt$ is invariant under the following space-time transformation
\begin{equation}
\la{bN1} t\longrightarrow T,\quad \bm{q}\longrightarrow \bm{Q}\,.
\end{equation}
In other words, we have
\bea\la{bN2}
\int L(t, \bm{q}, \dot{\bm{q}}) dt=\int L(T, \bm{Q}, \bm{\acute{Q}}) dT,
\eea
where $\bm{\acute{Q}}\equiv d\bm{Q}/dT$. Then, Noether theorem claims that the invariance of the action under (\ref{bN1}) will lead to existence of certain motion constants which are called Noether charges. When the transformations are generated by Killing vectors, the corresponding charges are called Killing-Noether charges.

Let's consider an infinitely small transformation, we write $T$ and $\bm{Q}$ in Eq.~\eqref{bN1} as follows,
\begin{eqnarray}\label{b34+1}
&& T=T(t,\bm{q},\dot{\bm{q}},\epsilon),\\
\label{b35}&& \bm{Q}=\bm{Q}(t,\bm{q},\dot{\bm{q}},\epsilon)\,,
\end{eqnarray}
where $\epsilon$ is an infinitesimal parameter being independent of the spacetime coordinates, and the following conditions hold,
\begin{eqnarray}\label{b36}
&& ~~  T|_{\epsilon=0}=t,\\
\label{b37}&&~~\bm{Q}|_{\epsilon=0}=\bm{q}\,.
\end{eqnarray}
The function $\bm{\acute{Q}}$ in the right-handed side of Eq.~\eqref{bN2} is then
\begin{eqnarray}\label{b38}
\bm{\acute{Q}}(t,\bm{q},\dot{\bm{q}},\ddot{\bm{q}},\epsilon)\equiv {d\bm{Q}\over d T}
={d\bm{Q}/dt\over dT/dt}
= {\dot{\bm{Q}}\over \dot{T}}=
{\dot{\bm{Q}}(t,\bm{q},\dot{\bm{q}},\ddot{\bm{q}},\epsilon)\over \dot{T}(t,\bm{q},\dot{\bm{q}},\ddot{\bm{q}},\epsilon)}\,.
\end{eqnarray}
We can also rewrite Eq.~\eqref{bN2} as
\begin{equation}\label{baa1}
\int [L(T,\bm{Q},\bm{\acute{Q}})\dot{T}-L(t, \bm{q},\dot{\bm{q}})]dt=0,
\end{equation}
then it can be proved that the following parameter is a motion integral constant~\cite{Noether,Arnold,yan2}:
\begin{equation}\label{b40}
G\equiv L\zeta+\sum_i{\pa L(t,\bm{q},\dot{\bm{q}})\over \pa \dot{q}^i}(\eta^i-\dot{q}^i\zeta)
\end{equation}
where
\begin{equation}
\label{eq:b41_b42}\zeta=\left.{\pa T(t, \bm{q},\dot{\bm{q}},\epsilon)\over \pa \epsilon}\right|_{\epsilon=0}\,,\quad \eta^i=\left. \pa Q^i(t, \bm{q},\dot{\bm{q}},\epsilon)\over \pa \epsilon\right|_{\epsilon=0}\,.
\end{equation}
Namely the conserved quantity $G$ of Eq.~\eqref{b40} satisfies
\bea\la{bcs1}
\dot{G}=0\,.
\eea

\noindent {\bf (B) Killing-Noether Charges}

Based on Killing vector equations Eqs.~(\ref{b3-2}, \ref{b3-3}) it can be showed~\cite{Weinberg72} that the infinitesimal transformation $x^\mu\rightarrow x'^\mu=x^\mu+\epsilon \xi^\mu$ leaves the Beltrami metric intact, i.e.
\bea\la{bshh1}
B_{\mu\nu}(x)\rightarrow B'_{\mu\nu}(x')={\pa x^\alpha\over \pa x'^\mu}{\pa x^\beta\over \pa x'^\nu}B_{\alpha\beta}(x)=B_{\mu\nu}(x').
\eea
Then it is easy to check that the Landau-Lifshitz action in Eq.~\eqref{b-3} is invariant under this metric preserved transformation,
\begin{small}
\begin{equation}
\la{bshh2}S\equiv -mc\int \sqrt{B_{\mu\nu}(x) dx^\mu dx^\nu}\rightarrow S'\equiv -mc\int \sqrt{B'_{\mu\nu}(x') dx'^\mu dx'^\nu}=-mc\int \sqrt{B_{\mu\nu}(x') dx'^\mu dx'^\nu}=S.
\end{equation}
\end{small}
Therefore, using the expressions of ten independent Killing vectors in Eq.~\eqref{bshh3} and the Noether theorem of Eq.~\eqref{b40}, the ten conserved quantities for dS/AdS-mechanics can be calculated out analytically.

\begin{enumerate}
\item Energy

Taking the constants in the Killing vector to be: $b_{\mu\nu}=0,\;a_1=a_2=a_3=0,\;a_0=-c$,  noting $x^0=ct$, and substituting them into Eqs.~(\ref{bcc}, \ref{bshh3}, \ref{b33-1}), we obtain
\bea\la{bshh4}
&&t'=t+{\epsilon\over c}\xi^0=t-\epsilon\left(1-{\Lambda\over3}c^2t^2\right),\\
\la{shh5}&& x'^i=x^i\left(1+{\Lambda c^2 t\epsilon\over 3}\right).
\eea
Comparing Eq.~\eqref{b3-1} with Eq.~\eqref{bN1} further, we have
\bea\la{bshh4}
t'=T,~~~~~\bm{x}'=\bm{Q}.
\eea
Thus the parameters $\zeta$ and $\eta^i$ defined in Eq.~\eqref{eq:b41_b42} take the form
\begin{eqnarray}\label{b3-88}
\zeta=-1+{\Lambda c^2t^2\over 3},\qquad
\eta^i=x^i{\Lambda c^2t\over 3}.
\end{eqnarray}
The corresponding Noether charge denoted as $G_{a^0}$ is given by
\begin{eqnarray}\nn
&&G_{a^0}=L\left(-1+{\Lambda c^2t^2\over 3}\right)+\sum_{i=1}^{3}\left[x^i{\Lambda c^2t\over 3}-\dot{x}^i\left(-1+{\Lambda c^2t^2\over 3}\right)\right] \left[{m^2c^2 \over  L} \right.\\
\label{b3-94} &&\times \left. {-9\dot{x}^i+3\Lambda[-\bm{x}^2\dot{x}^i +(\bm{x}\cdot\dot{\bm{x}})x^i-c^2t(x^i-\dot{x}^it)] \over  [3+\Lambda(\bm{x}^2-c^2t^2)]^2} \right] .
\end{eqnarray}
Inserting the expression of $L$ in Eq.~\eqref{b-4} into this equation, through an analytical calculation, we obtain
\begin{eqnarray}\label{b3-95}
G_{a^0}\equiv E={mc^2\over \sqrt{1-{\dot{\bm{x}}^2\over c^2}+{\Lambda(\bm{x}\cdot \dot{\bm{x}})^2 -\Lambda\bm{x}^2\dot{\bm{x}}^2\over 3c^2}+{\Lambda(\bm{x}-\dot{\bm{x}}t)^2\over 3}}},
\end{eqnarray}
which is desired energy formula for dS/AdS-SR mechanics. Moreover, we would like to make two remarks as follows,

\begin{itemize}
\item Introducing dS/AdS-SR Lorentz factor
\bea\la{b3-96}
\Gamma\equiv {1\over \sqrt{1-{\dot{\bm{x}}^2\over c^2}+\Lambda\left[{(\bm{x}\cdot \dot{\bm{x}})^2 -\bm{x}^2\dot{\bm{x}}^2\over 3c^2}+{(\bm{x}-\dot{\bm{x}}t)^2\over 3}\right]}},
\eea
then the energy in Eq.~\eqref{b3-95} and the Lagrangian in Eq.~\eqref{b-4} can be compactly written as
\bea\la{b3-97}
&& E=mc^2\Gamma,\\
\la{b3-97+}&& L=-mc^2(\sigma\Gamma)^{-1},
\eea
where $\sigma$ is given in Eq.~\eqref{sol}. In the limit of $\Lambda\rightarrow 0$, we have $\Gamma\rightarrow \gamma\equiv (1-\dot{\bm{x}}^2/c^2)^{-1/2}$, where $\gamma$ is usual Lorentz contraction factor of common SR. The energy $E=mc^2\Gamma$ goes back to common SR's energy formula $E=mc^2\gamma$, $L$ back to common Lagrangian of SR in Eq.~\eqref{b2-5}. Therefore it is reasonable to identify the Noether-charge $G_{a^0}$ as energy.
\item From the equation of motion $\ddot{\bm{x}}=0$ given in Eq.~\eqref{b-2}, it is easy to check that the equality $\dot{\Gamma}=0$ is fulfilled. Consequently we have
\bea\la{3-98}
\dot{E}=mc^2\dot{\Gamma}=0\,.
\eea
The energy conservation is verified in the dS/AdS-SR mechanics. Noting even though the Lagrangian for dS/AdS-SR in Eq.~\eqref{b-4} is time dependent, the corresponding energy is still conserving. This is a non-trivial character of the dS/AdS-SR Lagrangian formalism.
\end{itemize}

\item Momentum

Choosing the constants in the Killing vector to be $b_{\mu\nu}=0,\;a_0=a_2=a_3=0,\;a_1=-1$, accordingly the spacetime transformation is of the form
\bea\la{btp2}
&&t'=t+\epsilon {\Lambda\over3}tx^1,\\
\la{xp2} &&x'^i=x^i+\epsilon\left(\delta^{i1}+{\Lambda\over3} x^ix^1\right),~~~~i=1,\;2,\;3\,,
\eea
which lead to
\begin{eqnarray}\label{b3-101}
\zeta={\Lambda\over 3}tx^1,\quad
\eta^1=1+{\Lambda(x^1)^2\over3},\quad \eta^2={\Lambda x^1x^2\over3},\quad \eta^3={\Lambda x^1x^3\over3}\,.
\end{eqnarray}
Then we can straightforwardly determine the Noether charge is
\begin{eqnarray}\label{b3-109}
G_{a^1}\equiv p^1={m\dot{x}^1\over \sqrt{1-{\dot{\bm{x}}^2\over c^2}+\Lambda\left({(\bm{x}\cdot \dot{\bm{x}})^2 -\bm{x}^2\dot{\bm{x}}^2\over 3c^2}+{(\bm{x}-\dot{\bm{x}}t)^2\over 3}\right)}}\,.
\end{eqnarray}
Similarly, setting $a_i=-1\;(i=2\;\rm{or}\;3)$, and other parameters $\{b_{\mu\nu},\;a_i\}$ in the Killing vector (\ref{bshh3}) vanish, the resulting conserved quantity is fixed to be
\begin{equation}
\label{b33-86} G_{a^i}\equiv p^i={m\dot{x}^i\over \sqrt{1-{\dot{\bm{x}}^2\over c^2}+\Lambda\left[{(\bm{x}\cdot \dot{\bm{x}})^2 -\bm{x}^2\dot{\bm{x}}^2\over 3c^2}+{(\bm{x}-\dot{\bm{x}}t)^2\over 3}\right]}}=m\dot{x}^i\Gamma\,.
\end{equation}
Noting $\ddot{\bm{x}}=0$ and $\dot{\Gamma}=0$, the momentum conservation law also holds in dS/AdS-SR mechanics,
\bea\la{b33-87}
\dot{p}^i=0,~~{\rm{or}}~~\bm{\dot{p}}=0\,.
\eea

\item Lorentz boost

In the same fashion as previous cases, taking constants in the Killing vector to be: $b_{10}=1$ and other $b_{\mu\nu}=0$, $a_0=a_1=a_2=a_3=0$, we have
\begin{eqnarray}\nonumber
&& t'=t-{\epsilon x^1\over c},\\
\label{b3-111}&&x'^1=x^1-\epsilon ct,\quad x'^2=x^2,\quad x'^3=x^3,
\end{eqnarray}
and
\begin{eqnarray}\label{b33-101}
\zeta={-x^1\over c},\quad
\eta^1=-ct,\quad \eta^2=\eta^3=0.
\end{eqnarray}
The conserved quantity for this symmetry is given by
\begin{eqnarray}\label{b33-109}
G_{b_{10}}\equiv K^1={mc(x^1-t\dot{x}^1)\over \sqrt{1-{\dot{\bm{x}}^2\over c^2}+\Lambda\left[{(\bm{x}\cdot \dot{\bm{x}})^2 -\bm{x}^2\dot{\bm{x}}^2\over 3c^2}+{(\bm{x}-\dot{\bm{x}}t)^2\over 3}\right]}}\,.
\end{eqnarray}
Similarly for the case of $b_{i0}=1\;(i=2, 3)$ and other parameters in the Kiling vector vanishing, we find the corresponding Lorentz boost Noether charge takes the form
\begin{equation}
\label{b333-86} G_{b_{i0}}\equiv K^i={mc(x^i-t\dot{x}^i)\over \sqrt{1-{\dot{\bm{x}}^2\over c^2}+\Lambda\left[{(\bm{x}\cdot \dot{\bm{x}})^2 -\bm{x}^2\dot{\bm{x}}^2\over 3c^2}+{(\bm{x}-\dot{\bm{x}}t)^2\over 3}\right]}}=mc(x^i-t\dot{x}^i)\Gamma\,.
\end{equation}
Considering $\ddot{\bm{x}}=0$ and $\dot{\Gamma}=0$, we can easily checked that $K^i$ is really conserved in the dS/AdS-SR mechanics,
\bea\la{b333-87}
\dot{K}^i=0,~~{\rm{or}}~~\bm{\dot{K}}=0\,.
\eea

\item Angular momentum:

Finally, we derive the angular momentum of dS/AdS-SR mechanics. Taking $b_{12}=-1$, other $b_{\mu\nu}=0$ and $a_0=a_1=a_2=a_3=0$ in the Killing vector, we obtain
\begin{eqnarray}\label{b4-111}
t'=t,\quad x'^1=x^1-\epsilon x^2,\quad x'^2=x^2+\epsilon x^1,\quad x'^3=x^3,
\end{eqnarray}
and
\begin{eqnarray}\label{b4-101}
\zeta=0,\quad \eta^1=-x^2,\quad \eta^2=x^1,\quad \eta^3=0\,.
\end{eqnarray}
The conserved quantity is determined to be
\begin{eqnarray}\label{b4-109}
G_{b_{12}}\equiv L^3={m(x^1\dot{x}^2-x^2\dot{x}^1)\over \sqrt{1-{\dot{\bm{x}}^2\over c^2}+\Lambda\left[{(\bm{x}\cdot \dot{\bm{x}})^2 -\bm{x}^2\dot{\bm{x}}^2\over 3c^2}+{(\bm{x}-\dot{\bm{x}}t)^2\over 3}\right]}}\,.
\end{eqnarray}
For the choices of $b_{23}=-1$ and $b_{13}=-1$, the resulting Noether charges $G_{b_{23}}\equiv L^1=$ and $G_{b_{13}}\equiv -L^2$ can be calculated as follows
\begin{equation}
\label{b4-86}L^i={m\epsilon^{ijk}x^j\dot{x}^k\over \sqrt{1-{\dot{\bm{x}}^2\over c^2}+\Lambda\left[{(\bm{x}\cdot \dot{\bm{x}})^2 -\bm{x}^2\dot{\bm{x}}^2\over 3c^2}+{(\bm{x}-\dot{\bm{x}}t)^2\over 3}\right]}}=m\epsilon^{ijk}x^j\dot{x}^k\Gamma\,,
\end{equation}
where $\epsilon^{ijk}$ is the totally antisymmetric Levi-Civita symbol.
It is easy to verify that the angular momentum conservation law holds in dS/AdS-SR mechanics:
\bea\la{4-87}
\dot{L}^i=0,~~{\rm{or}}~~\bm{\dot{L}}=0\,.
\eea
\end{enumerate}
So far all the ten independent conserved Killing-Noether charges $\{E,\;\bm{p},\;\bm{K},\;\bm{L}\}$ have been found out. Comparing them with the corresponding results in~\cite{yan1} it is found that the Killing-Noether charges and the Noether charges deduced from the classical domain are exactly the same. Since the essential correctness of the classical domain method is less known in the community, our calculations in above are meaningful and useful for trusting in that method.

Existence of ten independent conserved Noether charges indicates also that the 4-dimension Beltrami spacetime has the maximal symmetry, and metric $B_{\mu\nu}(x)$ satisfies the $3^{\rm rd}$ requirement of basic metric for SR.

\section{Summary and discussion}
We show in this paper that when non-zero ECC (Einstein Cosmological Constant) $\Lambda$ emerges as a universal parameter in the Einstein equation, the Minkowski spacetime metric $\eta_{\mu\nu}$ of the common Special Relativity  is no longer a solution to the vacuum Einstein equation. This is a challenging puzzle in the relativity theories. The basic features for $\eta_{\mu\nu}$ are as follows: (i) It is the solution of vacuum Einstein equation with $\Lambda=0$; (ii) The inertial motion law of a free particle holds true in the Minkowski spacetime, hence we call $\eta_{\mu\nu}$ inertial metric; (iii) It has maximal spacetime symmetry. In order to understand the puzzle mentioned above, we start from Ref.~\cite{Lu74}. In Ref.~\cite{Lu74}, another inertial metric $g_{\mu\nu}(\lambda, x)$ with a parameter $\lambda$ were found by a miracle, and it was called {\it the classical domain metric} originally in~\cite{Lu74}. In the present paper, we have pursued this metric from two sides as follows:
\begin{enumerate}
\item Firstly, we successfully proved that when $\lambda=\Lambda/3$, the classical domain metric $g_{\mu\nu}(\lambda, x)$ satisfies the vacuum Einstein equation with $\Lambda\neq 0$, and named it Beltrami metric, i.e., $B_{\mu\nu}(x)=g_{\mu\nu}(\lambda,x)|_{(\lambda=\Lambda/3)}$. Thus, $B_{\mu\nu}(x)$ could be qualified to be the basic metrics of the dS/AdS-SR, if $B_{\mu\nu}(x)$ had maximal spacetime symmetry. Discussing the spacetime symmetry and the relevant physics is the main motivation of this work.

\item Secondly, therefore, we pay great attention to the Killing equations for $B_{\mu\nu}(x)$ and their general solutions. From Killing vector theory we gave the corresponding explicit expressions of the Killing equations which are system of ten simultaneous partial differential equations of first order. The general solutions for these Killing equations were obtained, and all ten independent Killing vectors were revealed explicitly. Such results confirm that the Beltrami metric $B_{\mu\nu}(x)$ has maximal spacetime symmetry. Since Killing vectors are the generators of the transformations preserving metric, the ten Killing-Noether charges should exist. The explicit form of these Noether charges have been calculated out. The results are just the energy, momenta, Lorentz boost and angular momentum $\{E,\;\bm{p},\;\bm{K},\;\bm{L}\}$ in SR-theory with $\Lambda\neq 0$.

\end{enumerate}

The pioneer work on dS/AdS-SR~\cite{Lu74} was based on the unpopular classical domain method. The present paper reformulates the theory of dS/AdS-SR by means of Killing vector geometric theory. Moreover, the study about the effects of vacuum with non-zero Einstein Cosmologic Constant is essential.

\begin{center} {\bf ACKNOWLEDGMENTS}
\end{center}
{The author thanks professor Gui-Jun Ding for stimulation discussions.  This work is partly supported by the National Nature Science Foundation of China numbered 11375169.}
\medskip



\begin{thebibliography}{99}

\bibitem{Dirac35} Dirac,P.A.M., The Electron Wave Function In De Sitter Space,  Annals of Mathmatics, 1935, {\bf 36}, No.3: 657-669.
\bibitem{Lu74} Lu Q K, Zou Z L, Guo H Y, Kinematics and cosmologic red shift phenomena in classical domain space-time (in Chinese). Acta Physica Sinica, 1974, 29: 225-233.
\bibitem{yan2} Yan M L. De Sitter Invariant Special Relativity. World Scintific Publising, Singapore, 2015. 262

\bibitem{Riess}  Riess A G,  et al.  Observational evidence from supernovae for an accelerating universe and a cosmological constant.  Astro. J. 1998, 116: 1009-1038
\bibitem{Perl} Perlmutter S ,  et al. Measurements of Omega and Lambda from 42 high redshift supernovae.  Astro. J.  1999, 517: 565-586  [astro-ph/9812133].
\bibitem{review} Peebles P J E, Ratra B, The cosmological constant and dark energy. Rev. Mod. Phys. 2003, 75: 559-606

\bibitem{review1} Padmanabhan T, 	
Cosmological constant: The Weight of the vacuum.
Phys. Rep. 2003, 380: 235-320.

\bibitem{yan3} Yan M L, Hu S, Huang W, et al. On determination of the geometric cosmological constant from the OPERA experiment of superluminal neutrinos. Modern Physics Letters A,
2011,  27: 1250041. arXiv:1112.6217 [hep-ph].

\bibitem{Guo2} Tian Y, Guo H Y, Huang C G et al. Mechanics and Newton-Cartan-like gravity on the Newton-Hooke space-time.
 Phys. Rev. 2005, D71: 044030

\bibitem{Guo2-1} Guo H Y, Zhou B, Tian Y, et al. The Triality of Conformal Extensions of Three Kinds of Special Relativity. Phys. Rev. 2007, D75: 026006
\bibitem{Guo2-2} Chang Z, Chen S X, Guan C B, et al. Cosmic ray threshold in an asymptotically dS spacetime.
Phys.Rev. 2005, D71: 103007

\bibitem{Guo1} Guo H Y, Huang C G, Xu Z, et al. On special relativity with cosmological constant. Phys. Lett. 2004, A331: 1-7

\bibitem{Guo1-1} Guo H Y, Huang C G, Xu Z, et al. On Beltrami Model of de Sitter Spacetime.
Mod. Phys. Lett. 2004, A19: 1701-1710
\bibitem{Guo1-2} Guo H Y, Huang C G, Xu Z, et al. Three kinds of special relativity via inverse Wick rotation.
 Chin. Phys. Lett. 2005, 22:  2477-2480.  hep-th/0405137
\bibitem{Lu05} Lu Q K, Heisenberg Group and Energy-Momentum Conservative Law in de-Sitter Spaces. Commu. Theor. Phys. 2005,  44: 389-392
\bibitem{yan1}
 Yan M L,  Xiao N C,  Huang W, et al. Hamiltonian Formalism of de-Sitter Invariant Special Relativity. Commu. Theor. Phys. 2007, 48: 27-36
 hep-th/0512319.

\bibitem{cxy} Chen S X, Xiao N C, Yan M L, Variation of the fine-structure constant from the de Sitter invariant special relativity. Chinese Phys. 2008, C32: 612-616


\bibitem{fy} Feng S S, Yan M L. Implication of Spatial and Temporal Variations of the Fine-Structure Constant.
Int J Theor Phys, 2016, 55: 1049-1083.
\bibitem{zhao} Zhao W, Santos L.  Preferred axis in cosmology, The Universe, 2015, no.3: 9-33 (Invited review), arXiv:1604.05484[astro-ph.CO]
\bibitem{ru} Tretyakova D A, Seeking for the observational manifestation of de Sitter Relativity. arXiv: 1604.00809[hep-ph]




\bibitem{yan4} Yan M L, One Electron Atom in Special Relativity with de Sitter Space-Time Symmetry. Commun. Theor. Phys. 2012, 57: 930-952
\bibitem{yan5} Yan M L, One Electron Atom in Special Relativity with de Sitter Space-Time Symmetry (II): ¡ª Higher
Order Contributions.
Commun. Theor. Phys. 2014, 62: 189-195


\bibitem{Landau}  Landau L D and Lifshits E M,  The Classical
Theory of Fields. Fourth revised English edition, Beijing:
Pergamon press, 1994. 402



\bibitem{hua2} Hua L K £¬Look K H (Lu Q K)£¬Theory of Harmonic Functions in Classical Domains. Scientia Sinica, 1959, 8: 1031-1094.

\bibitem{Weinberg72}  Weinberg S, Gravitation and Cosmology: Principles and Applications of the General Theory of Relativity. John Wiley $\&$ Sons, Inc. 1972. 657



\bibitem{Noether} Desloge E A, Classical Mechanics. John Wiley, New york, 1982. 991
\bibitem{Arnold}  Arnold V l, Mathematical Methods of Classical Mechanics, Springer, New York, 1989. 516

\end{thebibliography}
\end{document}